\documentclass[preprint2]{aastex63}
\usepackage{amsmath,amstext}
\usepackage{hyperref}
\usepackage[T1]{fontenc}
\usepackage[figure,figure*]{hypcap}
\usepackage{multirow}
\usepackage[T1]{fontenc} 
\usepackage[utf8]{inputenc} 

\newcommand{\UA}{Steward Observatory, The University of Arizona, 933 N.\ Cherry Ave, Tucson, AZ 85721, USA}
\newcommand{\PSUAA}{Department of Astronomy and Astrophysics, The Pennsylvania State University, 525 Davey Laboratory, University Park, PA 16802, USA}
\newcommand{\PSUCEHW}{Center for Exoplanets and Habitable Worlds, The Pennsylvania State University, 525 Davey Laboratory, University Park, PA 16802, USA}
\newcommand{\Princeton}{Department of Astrophysical Sciences, Princeton University, 4 Ivy Lane, Princeton, NJ 08540, USA}
\newcommand{\Macquarie}{Department of Physics and Astronomy, Macquarie University, Balaclava Road, North Ryde, NSW 2109, Australia}
\newcommand{\UCI}{Department of Physics and Astronomy, The University of California, Irvine, Irvine, CA 92697, USA}
\newcommand{\Carleton}{Department of Physics and Astronomy, Carleton College, One North College Street, Northfield, MN 55057, USA}
\newcommand{\NIST}{Time and Frequency Division, National Institute of Standards and Technology, 325 Broadway, Boulder, CO 80305, USA}
\newcommand{\CUBoulder}{Department of Physics, University of Colorado, 2000 Colorado Avenue, Boulder, CO 80309, USA}
\newcommand{\UT}{McDonald Observatory and Center for Planetary Systems Habitability, The University of Texas at Austin, Austin, TX 78730, USA}
\newcommand{\JPL}{Jet Propulsion Laboratory, California Institute of Technology, 4800 Oak Grove Drive, Pasadena, California 91109, USA}
\newcommand{\Caltech}{Department of Astronomy, California Institute of Technology, Pasadena, CA 91125, USA}
\newcommand{\UVA}{Department of Astronomy, University of Virginia, Charlottesville, VA 22904, USA}
\newcommand{\AFRL}{Space Vehicles Directorate, Air Force Research Laboratory, 3550 Aberdeen Ave. SE, Kirtland AFB, NM 87117, USA}
\newcommand{\PSUICS}{Institute for CyberScience, The Pennsylvania State University, University Park, PA, 16802, USA}
\newcommand{\NOAO}{NSF's National Optical-Infrared Astronomy Research Laboratory, Tucson, AZ 85719, USA}

\newcommand{\unit}[1]{\ensuremath{\, \mathrm{#1}}}
\newcommand\tess{TESS}
\newcommand\thistic{TOI-1899}
\newcommand\wrongtic{TIC 172370652}
\newcommand\kep{Kepler}
\newcommand\gaia{Gaia}

\shortauthors{Ca\~nas et al.}
\shorttitle{A warm Jupiter transiting an M dwarf}

\begin{document}
\title{A warm Jupiter transiting an M dwarf: A TESS single transit event confirmed with the Habitable-zone Planet Finder}
\correspondingauthor{Caleb I. Ca\~nas}
\email{canas@psu.edu}

\author[0000-0003-4835-0619]{Caleb I. Ca\~nas}
\altaffiliation{NASA Earth and Space Science Fellow}
\affiliation{\PSUAA}
\affiliation{\PSUCEHW}

\author[0000-0001-7409-5688]{Gudmundur Stefansson}
\altaffiliation{Henry Norris Russell Fellow}
\altaffiliation{NASA Earth and Space Science Fellow}
\affiliation{\Princeton}
\affiliation{\PSUAA}
\affiliation{\PSUCEHW}

\author[0000-0001-8401-4300]{Shubham Kanodia}
\affiliation{\PSUAA}
\affiliation{\PSUCEHW}

\author[0000-0001-9596-7983]{Suvrath Mahadevan}
\affil{\PSUAA}
\affil{\PSUCEHW}

\author[0000-0001-9662-3496]{William D. Cochran}
\affil{\UT}

\author[0000-0002-7714-6310]{Michael Endl}
\affil{\UT}

\author[0000-0003-0149-9678]{Paul Robertson}
\affil{\UCI}

\author[0000-0003-4384-7220]{Chad F.\ Bender}
\affil{\UA}

\author[0000-0001-8720-5612]{Joe P. Ninan}
\affil{\PSUAA}
\affil{\PSUCEHW}

\author[0000-0001-7708-2364]{Corey Beard}
\affil{\UCI}

\author[0000-0001-8342-7736]{Jack Lubin}
\affil{\UCI}

\author[0000-0002-5463-9980]{Arvind F. Gupta}
\affil{\PSUAA}
\affil{\PSUCEHW}

\author[0000-0002-0885-7215]{Mark E. Everett}
\affil{\NOAO}

\author[0000-0002-0048-2586]{Andrew Monson}
\affil{\PSUAA}
\affil{\PSUCEHW}

\author[0000-0002-4235-6369]{Robert F. Wilson}
\affil{\UVA}

\author[0000-0002-7871-085X]{Hannah M. Lewis}
\affil{\UVA}

\author{Mary Brewer}
\affil{\UVA}

\author[0000-0003-2025-3147]{Steven R. Majewski}
\affil{\UVA}

\author[0000-0003-1263-8637]{Leslie Hebb}
\affiliation{Department of Physics, Hobart and William Smith Colleges, 300 Pulteney Street, Geneva, NY 14456, USA}

\author[0000-0001-9677-1296]{Rebekah I. Dawson}
\affil{\PSUAA}
\affil{\PSUCEHW}

\author[0000-0002-2144-0764]{Scott A. Diddams}
\affil{\NIST}
\affil{\CUBoulder}

\author[0000-0001-6545-639X]{Eric B.\ Ford}
\affil{\PSUAA}
\affil{\PSUCEHW}
\affil{\PSUICS}

\author[0000-0002-0560-1433]{Connor Fredrick}
\affil{\NIST}
\affil{\CUBoulder}

\author[0000-0003-1312-9391]{Samuel Halverson}
\affil{\JPL}

\author[0000-0002-1664-3102]{Fred Hearty}
\affil{\PSUAA}
\affil{\PSUCEHW}

\author[0000-0002-9082-6337]{Andrea S.J. Lin}
\affil{\PSUAA}
\affil{\PSUCEHW}

\author[0000-0001-5000-1018]{Andrew J. Metcalf}
\affil{\AFRL}
\affil{\NIST}
\affil{\CUBoulder}

\author[0000-0002-2488-7123]{Jayadev Rajagopal}
\affil{\NOAO}

\author{Lawrence W. Ramsey}
\affil{\PSUAA}
\affil{\PSUCEHW}

\author[0000-0001-8127-5775]{Arpita Roy}
\altaffiliation{Robert A. Millikan Postdoctoral Fellow}
\affil{\Caltech}

\author[0000-0002-4046-987X]{Christian Schwab}
\affil{\Macquarie}

\author[0000-0002-4788-8858]{Ryan C. Terrien}
\affil{\Carleton}

\author[0000-0001-6160-5888]{Jason T.\ Wright}
\affil{\PSUAA}
\affil{\PSUCEHW}

\begin{abstract}
We confirm the planetary nature of a warm Jupiter transiting the early M dwarf TOI-1899, using a combination of available TESS photometry; high-precision, near-infrared spectroscopy with the Habitable-zone Planet Finder; and speckle and adaptive optics imaging. The data reveal a transiting companion on an $\sim29$-day orbit with a mass and radius of $0.66\pm0.07\ \mathrm{M_{J}}$ and $1.15_{-0.05}^{+0.04}\ \mathrm{R_{J}}$, respectively. The star TOI-1899 is the lowest-mass star known to host a transiting warm Jupiter, and we discuss the follow-up opportunities afforded by a warm ($\mathrm{T_{eq}}\sim362$ K) gas giant orbiting an M0 star. Our observations reveal that TOI-1899.01 is a puffy warm Jupiter, and we suggest additional transit observations to both refine the orbit and constrain the true dilution observed in \tess{}.
\end{abstract}

\keywords{planets and satellites: detection --- planetary systems --- stars: fundamental parameters}

\section{Introduction}
Close-orbiting Jupiter-sized exoplanets were one of the first types of exoplanets discovered. There is still no consensus as to the exact formation and migration mechanisms required to create this population. Predictions using the core accretion theory of planet formation suggest there is a low abundance of Jupiter-like planets orbiting M dwarfs \citep[e.g.,][]{Laughlin2004}. From radial velocity (RV) surveys, Jupiter-sized exoplanets are relatively rare in the Galaxy, and their occurrence rate decreases around the M dwarf population \citep[e.g.;][]{Endl2006,Johnson2010,Bonfils2013}. 

Of particular interest is the population of transiting warm Jupiters (WJs) that have periods spanning \(\sim10-200\) days because such systems allow us to probe migration pathways and test our understanding of planetary internal structures. The WJs are far enough from the host star that the stellar obliquity would remain unperturbed by tides raised on the star \citep[][but see also \cite{Li2016}]{Albrecht2012}, and any inflation in their radii should occur through delayed contraction and not via stellar flux-driven or tidal mechanisms \citep{Baraffe2014}. While ground-based surveys have been important in the detection and characterization of hot Jupiters with periods \(<10\) days, transiting WJs are challenging to discover from the ground. As of this writing, there are four known short-period (<10 days), transiting Jupiter-sized exoplanets orbiting M dwarfs: Kepler-45 b \citep{Johnson2012}, HATS-6 b \citep{Hartman2015}, NGTS-1 b \citep{Bayliss2018}, and HATS-71 b \citep{Bakos2020}. Some WJs orbiting M dwarfs have been detected through the RV method \citep[e.g.,][]{Marcy1998,Delfosse1998,Morales2019}, but none have been shown to transit.

In this paper, we confirm the planetary nature of a WJ transiting the M dwarf \thistic{} (TIC 172370679, \gaia{} DR2 2073530190996615424; $T=12.58$, $G_{RP}=12.59$). We characterize the system using adaptive optics (AO) imaging with the ShaneAO instrument \citep{srinath2014} on the 3 m Shane Telescope at Lick Observatory, speckle imaging with the NN-EXPLORE Exoplanet Stellar Speckle Imager \citep[NESSI;][]{Scott2018} instrument at the WIYN 3.5 m telescope, and precision near-infrared (NIR) RVs obtained with the Habitable-zone Planet Finder Spectrograph \citep[HPF;][]{mahadevan2012,mahadevan2014}. We derive stellar parameters for \thistic{} using our HPF spectra and use the HPF RVs to confirm the WJ nature of the transiting companion.

This paper is structured as follows. Section \ref{sec:statval} presents the photometric and imaging observations used to analyze the false-positive probability (FPP) of this planet, and Section \ref{sec:groundobs} presents the subsequent ground-based photometric and confirming spectroscopic observations of \thistic{}. Section \ref{sec:stellarpar} presents our best estimates of the stellar parameters, Section \ref{jointfit} describes the analysis of the photometry and velocimetry, Section \ref{sec:discussion} provides further discussion of the feasibility for future study of this system, and we conclude the paper in Section \ref{sec:summary} with a summary of our key results.

% ---------------
% ---------------
% ---------------
\section{Detection and Statistical Validation} \label{sec:statval}
\subsection{TESS Photometry}
\tess{} observed TIC 172370679 in Sectors 14 and 15 and has photometric data spanning 2019 July 18 through 2019 September 10. Given its single-transit nature, this target was not detected by the \tess{} science processing pipeline \citep[SPOC;][]{Jenkins2016} nor was it listed as a threshold-crossing event (TCE) by the \tess{} Science Office.\footnote{\url{https://archive.stsci.edu/tess/bulk_downloads/bulk_downloads_tce.html}} The \tess{} data validation statistics are similar to the \kep{} data validation statistics, and the classification as a TCE uses the multiple event statistic (MES), a value that gives the significance of a detection when the data is folded to the calculated orbital period \citep{Tenenbaum2013}. \kep{} adopted an MES threshold of \(7.1\sigma\) \citep{Jenkins2002} to ensure there was no more than one false-alarm detection during the entirety of the \kep{} mission when searching for an Earth-sized planet producing four transits around a 12th magnitude Sun-like star. Based on the data release notes for Sector 15,\footnote{\url{https://archive.stsci.edu/missions/tess/doc/tess_drn/tess_sector_15_drn21_v02.pdf}} \tess{} adopts an identical MES threshold, and any detection below this threshold, such as a single-transiting object, is rejected from further analysis. After submission of this manuscript, TIC 172370679 was identified as a community object of interest by citizen scientists in the Planet Hunters \tess{} project \citep{Eisner2020} and given the designation TOI-1899.\footnote{\url{http://www.planethunters.org/}}

\begin{figure*}[!ht]
\epsscale{1.15}
\plotone{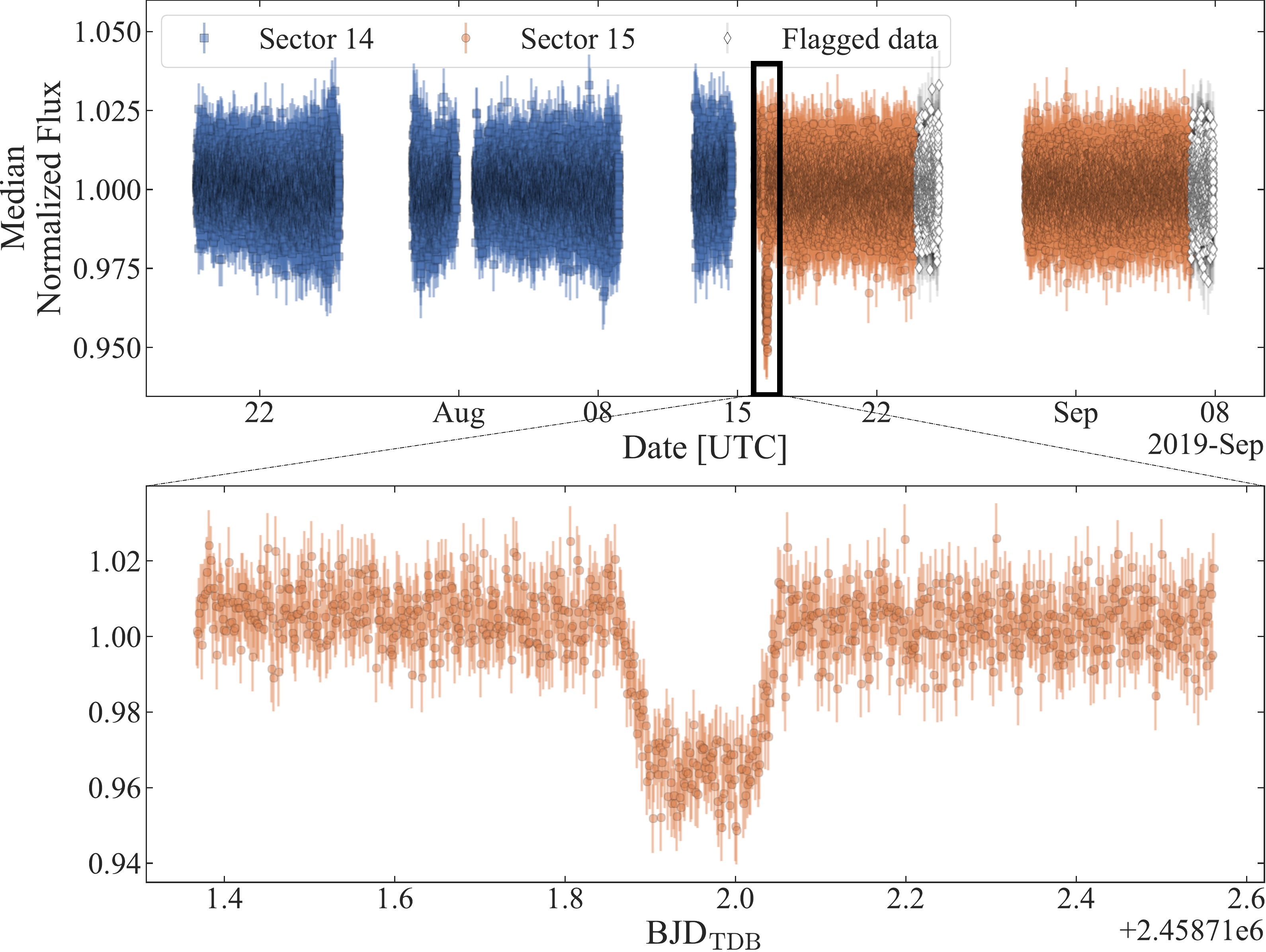}
\caption{The \tess{} photometry of \thistic{}. \textbf{Top} This panels presents the entire short-cadence PDCSAP \tess{} light curve for \thistic{}. Each sector is plotted with a different symbol and has been normalized by its respective median value. The white diamonds represent the data that were flagged by the \tess{} pipeline due to scattered-light contamination. The rectangle marks a region spanning \(\sim1.2\) days around the single-transit event observed in Sector 15. We use all of the \tess{} data that are not excluded by the quality flags for analysis in this work. \textbf{Bottom} The bottom panel is an enlarged version of the data contained in the rectangle. The observed transit is a single, \(\sim5\%\) flat-bottomed event with a duration of \(\sim 5\) hr.}
\label{fig:rawlc}
\end{figure*}

We identified \thistic{}.01 as a planetary candidate using a pipeline we developed to search for transiting candidates orbiting M dwarfs in the \tess{} short-cadence data. Our pipeline uses the \texttt{lightkurve} package \citep{LightkurveCollaboration2018} to detrend the data with a Savitzky-Golay filter and searches for transit events using the box least-squares algorithm \citep{Kovacs2002}. This target showed a single, \(\sim5\%\) flat-bottomed eclipsing event with a duration of \(\sim 5\) hr (Figure \ref{fig:rawlc}). Although only a single transit is visible in the \tess{} data, \thistic{} emerged as a promising WJ candidate for further follow-up observations due to the shape of the transit and the expected large RV semiamplitude of the planet. 

We searched for additional transits in the All-Sky Automated Survey for SuperNovae \citep[ASAS-SN;][]{Kochanek2017} and Zwicky Transient Facility \citep[ZTF;][]{Masci2019}. The ASAS-SN data have a mean cadence of one observation every 2 nights, but with the ASAS-SN photometric precision of \(\sim25\%\), only the transit of a binary star would be detected. The ZTF data have a mean cadence of one observation per night due to the simultaneous observations of the northern fields \citep{vanRoestel2019}. The ZTF has a photometric precision of \(\sim1\%\), but published observations are too sparse to sample the transit. Together, the ASAS-SN and ZTF data span over 1500 days but reveal no additional points during the observed \tess{} transit or any large-amplitude photometric variations that could be attributed to a close, bound stellar companion. 

For our subsequent analysis, we used the entire presearch data-conditioned time-series light curves \citep{Ricker2018} available at the Mikulski Archive for Space Telescopes (MAST) for Sectors 14 and 15. The data were processed by the SPOC, and the resulting light curve was corrected for dilution by known contaminating sources within the photometric aperture with a dilution factor of 0.756 (\texttt{CROWDSAP} in the SPOC light-curve file). We assumed the transit signal was superimposed on the photometric variability and that it could be detrended using a Gaussian process. We modeled the out-of-transit flux using the \texttt{celerite} package following the procedure in \cite{Foreman-Mackey2017} in which a simple function is constructed \citep[Equation (56) in][]{Foreman-Mackey2017} that mimics the properties of the quasiperiodic covariance function. No additional processing was performed on the light curve.

\subsection{Gaia Observations}
Given the large pixel size of \tess{}, dilution and other astrophysical false-positive scenarios must be evaluated prior to validation \citep[e.g.,][]{Sullivan2015}. To investigate the impact of background stars as a source of dilution, we searched the $11\times11$ \tess{} pixel grid centered on \thistic{} in \gaia{} DR2 \citep{GaiaCollaboration2018}. We use the \gaia{} $G_{RP}$ bandpass as an approximation to the \tess{} bandpass. Figure \hyperref[fig:tesspix]{2(a)} presents a ZTF $zr$ image overlaid with the \tess{} Sector 15 pixel grid and all stars identified in \gaia{} DR2 that have $|\Delta G_{RP}|<4$ when compared to \thistic{}. 

\gaia{} DR2 detects a total of 36 additional stars within the \tess{} aperture. The brightest stellar neighbor within this aperture, \wrongtic{} (\gaia{} DR2 2073530190996611200; $T=14.42$, $G_{RP}=14.38$), is \(17''\) away from \thistic{} and represents a flux ratio of \(\sim0.19\). \gaia{} DR2 reveals that \wrongtic{} is a giant star at a distance of \(2500\pm160\) pc and a radius of \(\sim4.5\unit{R_{\odot}}\). Given this size, if \wrongtic{} were the host star, the system would be an eclipsing binary. 

\subsection{Centroid and Aperture Analysis}
\begin{figure*}[!ht]
\epsscale{1.15}
\plotone{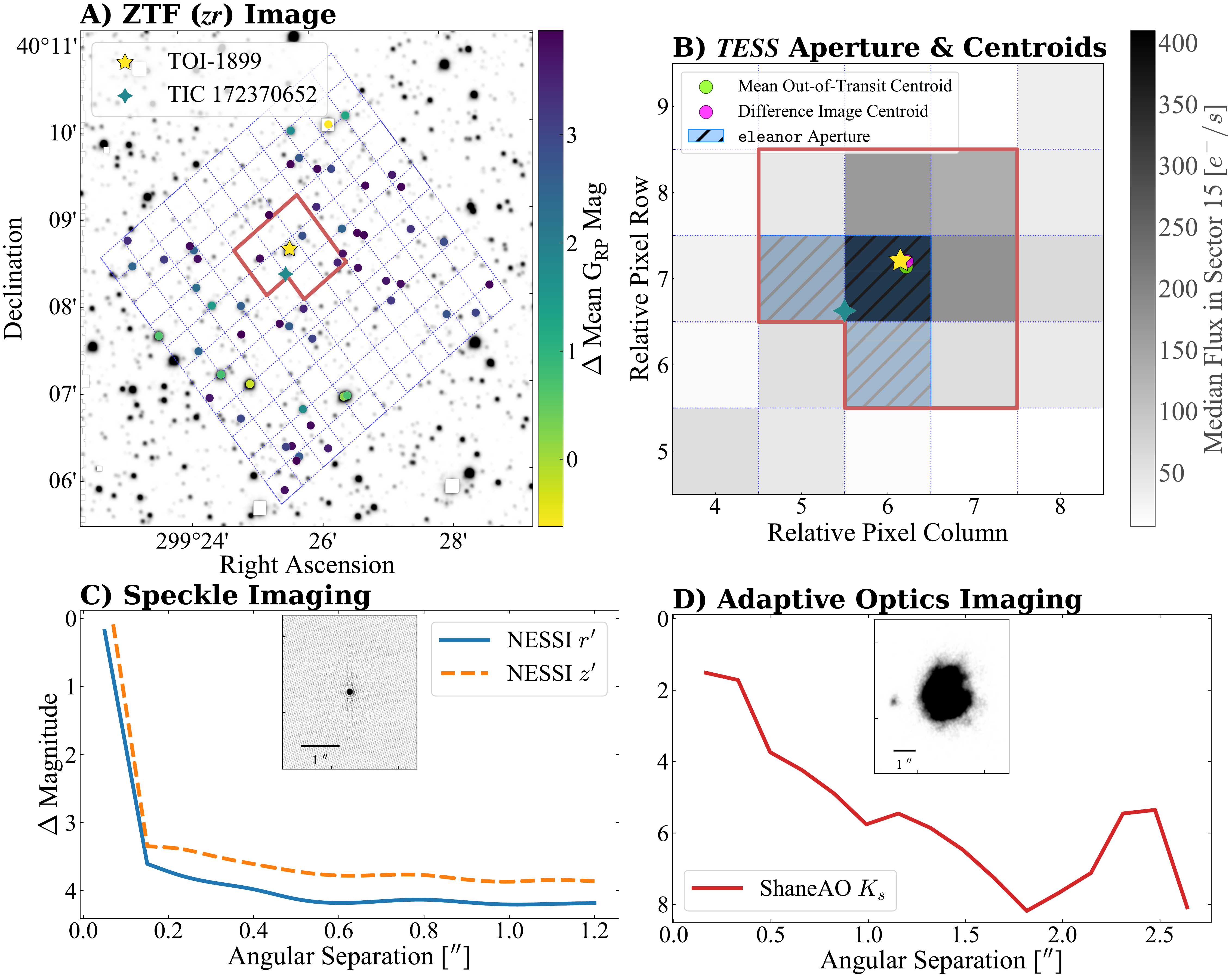}
\caption{Stellar  neighborhood around \thistic{}. \textbf{Panel (a)} shows the overlay of the \tess{} Sector 15 footprint on a ZTF \(zr\) image and highlights \thistic{} (star), the giant neighbor \wrongtic{} (diamond), and all stars with \(\Delta G_{RP}<4\) (circles). The \tess{} aperture is outlined in bold and contains three other stars with \(\Delta G_{RP}<4\), causing dilution of the transit. \textbf{Panel (b)} displays the region around the \tess{} aperture. Each pixel is colored to the median flux from Sector 15. The centroid does not significantly shift away from \thistic{} during transit, and this suggests that it is the host star. The hatched pixels denote the best aperture as determined from the full-frame images using \texttt{eleanor}. A light curve extracted with this aperture yields a transit depth identical to the short-cadence \tess{} data. We include the position of \thistic{} and \wrongtic{} for reference. \textbf{Panel (c)} displays the \(5\sigma\) contrast curve observed from NESSI in the Sloan \(r^\prime\) and \(z^\prime\) filters showing no bright companions within \(1.2''\) from the host star. The \(z^\prime\) image is shown as an inset. The horizontal line indicates the scale of 1\arcsec.  \textbf{Panel (d)} presents the \(5\sigma\) contrast curve observed from ShaneAO in the \(K_{s}\) filter with the detection of a \(\Delta K_{s}\approx5.5\) mag companion within \(2.4''\). The inset is the image from ShaneAO, and the horizontal line indicates the scale of 1\arcsec.}
\label{fig:tesspix}
\end{figure*}

To verify further that \wrongtic{} was only a source of dilution and not the source of the eclipsing event, we analyzed both the centroids and the aperture. We calculated the centroid during the \tess{} transit using Discovery and Vetting of Exoplanets \citep[\texttt{DAVE};][]{Kostov2019} to help distinguish between an eclipse occurring in the target system or in an unresolved background source. A significant centroid shift away from the purported target star during a transit is indicative of a false positive. \texttt{DAVE} is based on the methodology presented in \cite{Bryson2013} to compute the centroids by fitting a pixel response function model to the out-of-transit and difference images. The difference image is the difference between (i) the average of the flux before and after the transit and (ii) the flux during transit such that, in the difference image, the pixels containing the transit are regions of excess flux. The centroids are shown in Figure \hyperref[fig:tesspix]{2(b)}. Both the out-of-transit and in-transit centroids are located in the aperture pixel containing the most flux and are separated by 0.004 pixels (\(\sim 0.08\arcsec\)). This offset is more than 100 times smaller than the width of the point-spread function. The lack of a significant shift away from \thistic{} during transit is consistent with this star being the host star. 

We employed \texttt{eleanor} \citep{Feinstein2019} to probe which aperture is preferred in the \tess{} full-frame images. We used a segment of \(31\times31\) pixels in the calibrated full-frame images centered on \thistic{} to model the background and correct for systematics. Here \texttt{eleanor} derives light curves for various combinations of apertures and adopts the aperture that minimizes the combined differential photometric precision (CDPP) on the data when binned into 1 hr timescales. The CDPP was originally defined for \kep{} and is formally the rms of the photometric noise on transit timescales \citep{Jenkins2010}. Minimizing this metric ensures that sharp features on relatively short timescales, such as transits, are preserved. 

The preferred \texttt{eleanor} aperture is an L-shaped wedge centered on our star.  Figure \hyperref[fig:tesspix]{2(b)} presents the preferred \texttt{eleanor} aperture, which still includes the pixel containing the giant star \wrongtic{}. A light curve derived with this aperture from the full-frame images (using \texttt{psf\_flux} from \texttt{eleanor}) reveals a transit of identical depth to the dilution-corrected PDCSAP transit from the SPOC. Given the single-transiting nature of this object and the consistent depth, we opted to use the \tess{} short-cadence data for further analysis.

\subsection{Speckle Imaging}
To probe for binary companions or background objects, we performed speckle imaging using NESSI on the 3.5 m WIYN Telescope at KPNO on 2019 November 14. Due to the faintness of \thistic{}, the images were acquired in Sloan \(r^\prime\) and \(z^\prime\) instead of the narrower filters that NESSI traditionally uses. The images were reconstructed following the procedures outlined in \cite{Howell2011}. The NESSI contrast curves in both filters are shown in Figure \hyperref[fig:tesspix]{2(c)}, along with an inset of the \(z^\prime\) image. The NESSI data show no evidence of blending from a bright companion at separations of $0.1-1.2\arcsec$. 

\subsection{AO Imaging}
We performed high-contrast AO imaging of \thistic{} using the 3m Shane Telescope at Lick Observatory on 2019 November 10. The AO imaging was carried out using the upgraded ShARCS camera \citep{srinath2014} in the $K_s$ bandpass. We observed \thistic{} using a five-point dither pattern \citep[see, e.g.,][]{furlan2017}, imaging the star at the center of the detector and in each quadrant. We took images at four positions instead of the normal five because the limitations of the telescope motors prevented us from offsetting to one of the four standard off-center dither positions. Our experience is that sufficient sky subtraction can be performed with three or more of the five standard positions without a meaningful impact on the results.

Standard image processing, including flat-fielding, sky subtraction, and subpixel image alignment, was performed with custom Python software. We computed the variance in flux in a series of concentric annuli centered on the target star in the combined image. The resulting $5\sigma$ contrast curve is shown in Figure \hyperref[fig:tesspix]{2(b)}. From the images, we see that a faint ($\Delta K_s\approx5.5$) secondary companion is detectable at $\sim2.2\arcsec$. The amount of dilution attributable to this companion (\gaia{} DR2 2073530190984193280, TIC 1879763195; $T=18.63$) is negligible. These data show that there is no evidence of blending from a bright companion with up to $2.5\arcsec$ separation. 

\subsection{Statistical Validation}
We employed \texttt{VESPA} \citep{Morton2012} to conduct a false-positive analysis of \thistic{}.01. The algorithm validates a planet statistically by simulating and determining the likelihood of a range of astrophysical false-positive scenarios that include background eclipsing binaries (BEBs), eclipsing binaries, and hierarchical eclipsing binaries. The code generates a population for each false-positive scenario to calculate the likelihoods. 

For our analysis, we set Gaussian priors on the (i) Two Micron All Sky Survey (2MASS) \(JHK\) magnitudes \citep{Skrutskie2006}, (ii) Sloan Digital Sky Survey (SDSS) \(g^\prime r^\prime\) magnitudes from the AAVSO Photometric All-Sky Survey \citep[APASS;][]{Henden2015}, (iii) \gaia{} DR2 parallax, and (iv) host star surface gravity, temperature, and metallicity from the \tess{} Input Catalog \citep[TIC;][]{Stassun2019}, as well as a uniform prior on the visual extinction where the upper limit is determined using estimates of Galactic dust extinction by \cite{Green2019}. We set the maximum radius permissible for a BEB as the radius of the \tess{} aperture (\(48''\)). We constrain the maximum depth of the secondary transit as the rms of the light curve after excising the transit (\(<7700\) ppm). We include the ShaneAO and NESSI contrast curves shown in Figure \ref{fig:tesspix} as additional constraints applied to the BEB population during the \texttt{vespa} analysis. For this analysis, we adopted the period of \(P=30\) days, which we estimated by fitting the transit with a prior on the stellar density and assuming \thistic{}.01 was on a circular orbit (\(e=0\)). 

The WJ \thistic{}.01 has an FPP of \(0.004\). We note that \texttt{vespa} is a tool designed for the \kep{} mission that had pixels of \(\sim4''\) in size. With the \(\sim21''\) pixels of \tess{}, there will be blended stars in a given pixel. As such, we expect the FPP to be slightly underestimated for \tess{} photometry, particularly in crowded fields with known blends. For \thistic{}.01, our analysis reveals a marginally validated planet when adopting the threshold of FPP \(<1\%\) used in \cite{Morton2016}. The FPP was small enough to warrant subsequent spectroscopic observations. 

\section{Confirmation and Additional Observations} \label{sec:groundobs}
\subsection{High-resolution Doppler Spectroscopy}
We obtained 15 visits of \thistic{} using the HPF, a high-resolution ($R\sim55,000$), NIR (\(8080-12780\)\ \AA) spectrograph located at the 10m Hobby-Eberly Telescope (HET) in Texas \citep{mahadevan2012,mahadevan2014}. The HET is a fully queue-scheduled telescope with all observations executed in a queue by the HET resident astronomers \citep{shetrone2007}. The HPF is actively temperature-stabilized and achieves $\sim$1$\unit{mK}$ temperature stability long-term \citep{stefansson2016}. We use the algorithms in the tool \texttt{HxRGproc} for bias noise removal, nonlinearity correction, cosmic-ray correction, and slope/flux and variance image calculation \citep{Ninan2018} of the raw HPF data. We obtained two 945 s exposures per visit, except on the first visit, where we obtained only one exposure due to poor weather. This resulted in 29 spectra with a median signal-to-noise ratio (S/N) of 65 at $1000\unit{nm}$. While HPF has an NIR laser-frequency comb (LFC) calibrator that is shown to enable $\sim20\unit{cm\ s^{-1}}$ calibration precision and $1.53 \unit{m\ s^{-1}}$ RV precision on-sky \citep{metcalf2019}, we did not use the simultaneous LFC reference calibrator to minimize the impact of scattered LFC light in the target spectrum. We performed drift correction by extrapolating the wavelength solution from other LFC exposures from the night of the observations, as discussed in \cite{Stefansson2020}. This methodology enables precise wavelength calibration and drift correction up to $\sim30\unit{cm/s}$ per-observation, a value much smaller than our estimated per observation RV uncertainty for \thistic{} (at the $\sim$15$\unit{m\ s^{-1}}$ level).

The RVs are derived following the methodology described in \cite{Stefansson2020} using a modified version of the \texttt{SpEctrum Radial Velocity AnaLyser} pipeline \citep[\texttt{SERVAL};][]{zechmeister2018}. \texttt{SERVAL} employs the template-matching technique to derive RVs \citep[e.g.,][]{Anglada-Escude2012} by creating a master template from the observations to determine the Doppler shift for each individual spectrum by minimizing the \(\chi^2\) statistic. We generated the master template using all observed spectra while ignoring any telluric regions identified using a synthetic telluric-line mask generated from \texttt{telfit} \citep{Gullikson2014}, a Python wrapper to the Line-by-Line Radiative Transfer Model package \citep{clough2005}. We calculated the barycentric correction for each epoch using \texttt{barycorrpy}, the Python implementation \citep{Kanodia2018} of the algorithms from \cite{wright2014}. The observations are plotted in the top panel of Figure \hyperref[fig:rvs]{3}. Table \ref{tab:rvs} presents the derived RVs, the \(1\sigma\) uncertainties, and the S/N per pixel at 1000 nm for each epoch. 

\begin{deluxetable}{lrcc}
\tablecaption{RVs of \thistic{}. All observations have exposure times of 1890s unless otherwise indicated. \label{tab:rvs}}
\tablehead{\colhead{$\unit{BJD_{TDB}}$}  &  \colhead{RV}                   & \colhead{$\sigma$}  & \colhead{S/N} \\
           \colhead{}                    &  \colhead{$(\unit{m/s})$} & \colhead{$(\unit{m/s})$}  & \colhead{$@1000 \unit{nm}$} }
\startdata
 2458763.683421\(^a\) &   101.98 &   47.26 &   34 \\
 2458778.653989 &   -34.49 &   15.39 &   68 \\
 2458782.630853 &   -33.12 &   13.70 &   75 \\
 2458784.631621 &    -3.94 &   13.41 &   77 \\
 2458789.621115 &    53.50 &   14.82 &   71 \\
 2458793.603541 &    53.17 &   23.05 &   47 \\
 2458802.589058 &   -20.61 &   21.27 &   50 \\
 2458803.572612 &   -47.90 &   18.16 &   59 \\
 2458805.580779 &    -3.08 &   15.69 &   67 \\
 2458809.561619 &   -30.07 &   15.19 &   68 \\
 2458810.557411 &   -14.02 &   25.59 &   42 \\
 2458811.554959 &   -26.59 &   12.43 &   83 \\
 2458818.555145 &    67.27 &   32.03 &   36 \\
 2458819.542258 &    92.90 &   16.34 &   67 \\
 2458820.547178 &    89.73 &   16.54 &   64 \\
\enddata
\tablenotetext{a}{Exposure time is 945s.}
\end{deluxetable}
\subsection{Ground-based Photometry}
\subsubsection{HDI}
Once we obtained the first six RV measurements, it was clear that the data spanned one orbit of this system. We used a circular fit to the available HPF data to find the most probable transit times. To search for an additional transit, we observed \thistic{} each night between 2019 November 12 and November 15 with the Half-Degree Imager \citep[HDI;][]{Deliyannis2013} on the WIYN 0.9 m Telescope at KPNO. The HDI has a $4096 \times 4096 \unit{pixel}$ back-illuminated CCD with a $29.2\arcmin \times 29.2\arcmin$ field of view (FOV) at a plate scale of $0.425 \unit{\arcsec/pixel}$. All of the observations were performed slightly defocused and in the SDSS $z^\prime$ filter using the $1 \times 1$ binning mode and the four-amplifier readout mode.

We reduced the HDI observations using AstroImageJ \citep{Collins2017}, following the methodology described in \cite{stefansson2017} and \cite{stefansson2018}. For each night, we varied the radii of the software aperture and inner and outer background annuli in the reduction and adopted an object aperture radius of 10 pixels (4.25\arcsec), and inner and outer sky annuli of 15 pixels (6.38\arcsec) and 25 pixels (8.50\arcsec), respectively. This configuration resulted in the minimum scatter in the photometry. Figure \hyperref[fig:tesspix]{2(a)} presents the reduced HDI photometry, which shows no additional transits of \thistic{}.01. Subsequent observations with HPF better constrained the orbit of \thistic{} and clarified that our ground-based observations did not coincide with the expected mid-transit time.

\subsubsection{ARCSAT}
We observed \thistic{} on the night of 2019 November 13 using the SDSS $i^\prime$ filter using FlareCam on the Astrophysical Research Consortium Small Aperture Telescope (ARCSAT) located at Apache Point Observatory. ARCSAT, formerly known as the SDSS Photometric Telescope, is a 0.5 m telescope originally used to calibrate photometry for SDSS \citep{York2000,Tucker2006}. FlareCam is optimized for fast readout times and equipped with a 1024$\times$1024 back-illuminated CCD for enhanced sensitivity in the blue and near-UV with an $11.2\arcmin \times 11.2\arcmin$ FOV resulting in a pixel scale of $0.656 \unit{\arcsec/pixel}$ \citep{Hilton2011}.

The ARCSAT observations were carried out defocused with $2\times2$~binning, resulting in a pixel scale of 1.312 \arcsec/pixel. The data were bias- and dark-subtracted and flat-field corrected with use of the Python package \texttt{ccdproc} \citep{Craig2017}. We performed aperture photometry using the Python package \texttt{Photutils} \citep{Bradley2019}. We used an aperture radius of 5 pixels (6.5 \arcsec) and sky subtracted with an annulus having inner and outer radii of 7 pixels (9.1 \arcsec) and 11 pixels (14.3 \arcsec), respectively. This configuration minimized the scatter in the data while avoiding flux contamination from nearby sources in the chosen apertures. The ARCSAT data in Figure \hyperref[fig:rvs]{3(a)} are consistent with the HDI data and show no additional transit.

\section{Stellar Parameters} \label{sec:stellarpar}
\subsection{Spectroscopic Parameters}
We employed a modified version of the \texttt{SpecMatch-Emp} algorithm \citep{Yee2017} to characterize the properties of \thistic{} by comparing its highest-S/N spectrum to a library of high-resolution (\(R\sim55,000\)), high-quality (\(\unit{S/N}>100\)) HPF stellar spectra that have well-determined properties from \cite{Yee2017}. The modified HPF \texttt{SpecMatch-Emp} algorithm is described in \cite{Stefansson2020}. 

In brief, \texttt{SpecMatch-Emp} shifts the observed spectrum to the library wavelength scale, finds the best-matching library spectrum using \(\chi^{2}\) minimization, and uses a linear combination of the five best-matching spectra to synthesize a composite spectrum. We perform a cross-validation procedure where a spectrum from the library is removed, and we compare the recovered best-fit stellar parameter to its known library value. We repeat this comparison for the entire stellar library and adopt the standard deviation ($\sigma$) of the residuals between the recovered best-fit stellar parameters and the known library value as the uncertainty in each measurement ($\sigma_{T_{\mathrm{eff}}}$, $\sigma_{\mathrm{Fe/H}}$, and $\sigma_{\log g}$). 

As of this writing, the HPF \texttt{SpecMatch-Emp} library consists of 55 stars spanning the following parameter ranges: $3100 \unit{K} < T_{e} < 5000 \unit{K}$, $4.45<\log g < 5.12$, and $-0.5 < \mathrm{[Fe/H]} < 0.5$. Our comparisons use the wavelength region between $10460$ and $10570$ \AA\ because it is a region with minimal telluric contamination in the $Y$ band. The derived parameters for \thistic{} are $T_{\mathrm{eff}}=3925\pm 77 \unit{K}$, $\mathrm{[Fe/H]} = 0.20 \pm 0.13$ and $\log (g) = 4.68 \pm 0.05$. These values are comparable to the photoastrometric parameters derived with \texttt{StarHorse} \citep{Santiago2016,Queiroz2018}, a tool designed for Bayesian inference of stellar parameters and distances using data from spectroscopic surveys. The \texttt{StarHorse} values are $T_{\mathrm{eff}}=3945_{-37}^{+108} \unit{K}$, $\mathrm{Fe/H} = 0.19_{-0.17}^{+0.08}$ and $\log (g) = 4.65 \pm 0.01$. We adopt our \texttt{SpecMatch-Emp} parameters, as they are derived from spectra that also provide a reliable constraint on stellar metallicity. The derived spectroscopic parameters with their uncertainties are listed in Table \ref{tab:stellarparam}. Using our HPF spectra, we also place a formal constraint of $v \sin i_{*} < 2 \unit{km/s}$.

\subsection{Spectral Classification}
The best-matching library spectrum across all HPF spectral orders analyzed is GJ 1172, an M0 star \citep{Gaidos2014}. To confirm this spectral subtype, we used the catalog of M-type stars identified by the Large Sky Area Multi-Object Fibre Spectroscopic Telescope (LAMOST) collaboration \citep{Zhong2019}. LAMOST is a 4m telescope equipped with 4000 fibers distributed over a 5\degr\ FOV that is capable of acquiring spectra in the optical band (3700-9000\AA) at a resolution \(R\approx1800\) with a limiting magnitude of SDSS \(r^\prime=19\) mag \citep{Cui2012}.

The LAMOST stellar classification pipeline uses stellar templates to identify molecular absorption features (e.g., CaH, TiO) that are typical for M-type stars. To be classified as M dwarfs, targets must have (i) a mean S/N$>5$, (ii) a best-matching template that is an M type, and (iii) the spectral indices of the absorption features must be located in the M-type stellar regime identified in \cite{Zhong2019} (0 < TiO5 < 1.2 and 0.6< CaH2+CaH3 < 2.4). 

While the metallicities of the M dwarfs are not provided, the LAMOST M dwarf catalog does include a coarse indicator of metallicity, \(\zeta\). The value of this parameter is based on the strength of the TiO5, CaH2, and CaH3 molecular bands and quantifies the weakening of the TiO band strength due to metallicity effects \citep{Lepine2007}. \cite{Mann2013} tested the \(\zeta\) parameter with their sample and found that it correlates with [Fe/H] for supersolar metallicities but it does not necessarily correlate in metal-poor M dwarfs.

The proximity of \thistic{} to the original \kep{} field resulted in two observations with LAMOST as part of their \kep{} survey \citep{Zong2018}. From each observation, the spectral indices are consistent with an M0 classification. The mean value of \(\zeta=1.326\pm0.003\) suggests that this is a metal-rich M dwarf. The LAMOST classification as a metal-rich M0 dwarf is in agreement with our classification from \texttt{SpecMatch-Emp}.

\subsection{Model-dependent Stellar Parameters}
We used the \texttt{EXOFASTv2} analysis package \citep{Eastman2019} to model the spectral energy distribution (SED) and derive the stellar parameters using MIST stellar models \citep{Choi2016,Dotter2016}. We assumed Gaussian priors using the (i) 2MASS \(JHK\) magnitudes; (ii) SDSS \(g^\prime i^\prime\) and Johnson \(B\) magnitudes from APASS; (iii) Wide-field Infrared Survey Explorer magnitudes \citep{Wright2010}; (iv) host star surface gravity, temperature, and metallicity derived with \texttt{SpecMatch-Emp}; and (v) distance estimate from \cite{Bailer-Jones2018}. We adopt a uniform prior for the visual extinction where the upper limit is determined from estimates of Galactic dust by \cite{Green2019} (Bayestar19) calculated at the distance determined by \cite{Bailer-Jones2018}. We adopt the \(R_{v}=3.1\) reddening law from \cite{Fitzpatrick1999} to convert the Bayestar19 extinction to a visual magnitude extinction. The stellar priors and derived stellar parameters with their uncertainties are listed in Table \ref{tab:stellarparam}.

\startlongtable
\begin{deluxetable*}{lccc}
\tablecaption{Summary of Stellar Parameters. \label{tab:stellarparam}}
\tablehead{\colhead{~~~Parameter}&  \colhead{Description}&
\colhead{Value}&
\colhead{Reference}}
\startdata
\multicolumn{4}{l}{\hspace{-0.2cm} Main identifiers:}  \\
~~~TIC &  \(\cdots\)  & 172370679 & Stassun \\
~~~2MASS & \(\cdots\) & 19574239+4008357 & 2MASS \\
~~~Gaia DR2 & \(\cdots\) & 2073530190996615424 & Gaia \\
\multicolumn{4}{l}{\hspace{-0.2cm} Equatorial Coordinates, Proper Motion and Spectral Type:} \\
~~~$\alpha_{\mathrm{J2000}}$ &  Right Ascension (RA) & 19:57:42.44 & Gaia \\
~~~$\delta_{\mathrm{J2000}}$ &  Declination (Dec) & 40:08:36.05 & Gaia \\
~~~$\mu_{\alpha}$ &  Proper motion (RA, \unit{mas/yr}) & $35.427\pm0.025$ & Gaia \\
~~~$\mu_{\delta}$ &  Proper motion (Dec, \unit{mas/yr}) & $18.828\pm0.029$ & Gaia  \\
~~~$D$ & Dilution factor of \tess{} photometry &
$0.757$ & SPOC\\
~~~$d$ &  Distance in pc  & $128.4\pm0.3$ & Bailer-Jones\\
~~~\(A_{V,max}\) & Maximum visual extinction & $0.02$ & Green\\
~~~Spectral Type & \(\cdots\) & M0 & LAMOST \\
\multicolumn{4}{l}{\hspace{-0.2cm} Optical and near-infrared magnitudes:}  \\
~~~$B$ & Johnson B mag & $ 15.898\pm0.029$ & APASS\\
~~~$g^{\prime}$ &  Sloan $g^{\prime}$ mag  & $15.115\pm0.054$ & APASS\\
~~~$r^{\prime}$ &  Sloan $r^{\prime}$ mag  & $13.728\pm0.040$ & APASS \\
~~~$T$  & \tess{} magnitude & $12.582\pm0.007$  & Stassun \\
~~~$J$ & $J$ mag & $11.342\pm0.022$ & 2MASS\\
~~~$H$ & $H$ mag & $10.666\pm0.022$ & 2MASS\\
~~~$K_s$ & $K_s$ mag & $10.509\pm0.018$ & 2MASS\\
~~~$W1$ &  WISE1 mag & $10.412\pm0.022$ & WISE\\
~~~$W2$ &  WISE2 mag & $10.460\pm0.021$ & WISE\\
~~~$W3$ &  WISE3 mag & $10.312\pm0.045$ & WISE\\
\multicolumn{4}{l}{\hspace{-0.2cm} Spectroscopic Parameters$^a$:}\\
~~~$T_{e}$ &  Effective temperature in \unit{K} & $3925\pm77$& This work\\
~~~$\mathrm{[Fe/H]}$ &  Metallicity in dex & $0.20\pm0.13$ & This work\\
~~~$\log(g)$ & Surface gravity in cgs units & $4.68\pm0.05$ & This work\\
\multicolumn{4}{l}{\hspace{-0.2cm} Model-Dependent Stellar SED and Isochrone fit Parameters$^b$:}\\
~~~$T_{e}$ &  Effective temperature in \unit{K} & $3841_{-45}^{+54}$ & This work\\
~~~$\mathrm{[Fe/H]}$ & Metallicity in dex & $0.31_{-0.12}^{+0.11}$ & This work \\
~~~$\log(g)$ &  Surface gravity in cgs units & $4.669_{-0.022}^{+0.025}$ & This work \\
~~~$M_*$ &  Mass in $M_{\odot}$ & $0.627_{-0.028}^{+0.026}$ & This work\\
~~~$R_*$ &  Radius in $R_{\odot}$ & $0.607_{-0.023}^{+0.019}$ & This work\\
~~~$\rho_*$ &  Density in $\unit{g/cm^{3}}$ & $3.95_{-0.29}^{+0.37}$ & This work\\
~~~Age & Age in Gyrs & $7.4_{-4.6}^{+4.4}$ & This work\\
~~~$A_v$ & Visual extinction in mag & $0.010\pm0.007$ & This work\\
\multicolumn{4}{l}{\hspace{-0.2cm} Other Stellar Parameters:}           \\
~~~$v \sin i_*$ &  Rotational velocity in \unit{km/s}  & $<2$ & This work\\
~~~$RV$ &  Radial velocity in \unit{km/s} & $-28.95 \pm 0.07$ & This work\\
\enddata
\tablenotetext{}{References are: Stassun \citep{Stassun2018}, 2MASS \citep{Cutri2003}, Gaia \citep{GaiaCollaboration2018}, SPOC \citep{Jenkins2016}, Bailer-Jones \citep{Bailer-Jones2018}, Green \citep{Green2019}, LAMOST \citep{Zhong2019}, APASS \citep{Henden2015}, WISE \citep{Wright2010}}
\tablenotetext{a}{Derived using our modified \texttt{SpecMatch-Emp} algorithm.}
\tablenotetext{b}{\texttt{EXOFASTv2} derived values using MIST isochrones with the \gaia{} parallax and spectroscopic parameters in $a$) as priors.}
\end{deluxetable*}

\section{Data Analysis}\label{jointfit}
We employ the \texttt{juliet} analysis package \citep{Espinoza2019} to jointly model the photometry and velocimetry. The \texttt{juliet} package utilizes publicly available tools to model the photometry \citep[\texttt{batman};][]{Kreidberg2015} and velocimetry \citep[\texttt{radvel};][]{Fulton2018} and performs the parameter estimation using the importance nest-sampling algorithm \texttt{MultiNest} \citep{Feroz2013,Buchner2014}. The photometric model is based on the analytical formalism of \cite{Mandel2002} for a planetary transit and assumes a quadratic limb-darkening law in which the limb-darkening parameters are sampled using the $q_1$ and $q_2$ parameterization from \cite{Kipping2013}. We used the PDCSAP flux, which already corrects for dilution, so our photometric model does not include any additional dilution factor. We also set a prior on the stellar density using the value determined from our \texttt{EXOFASTv2} SED fit. The RV model is a standard Keplerian model. Both the photometric and RV models include a simple white-noise model in the form of a jitter term that is added in quadrature to the error bars of each data set.

\begin{figure*}[!ht]
\epsscale{1.15}
\plotone{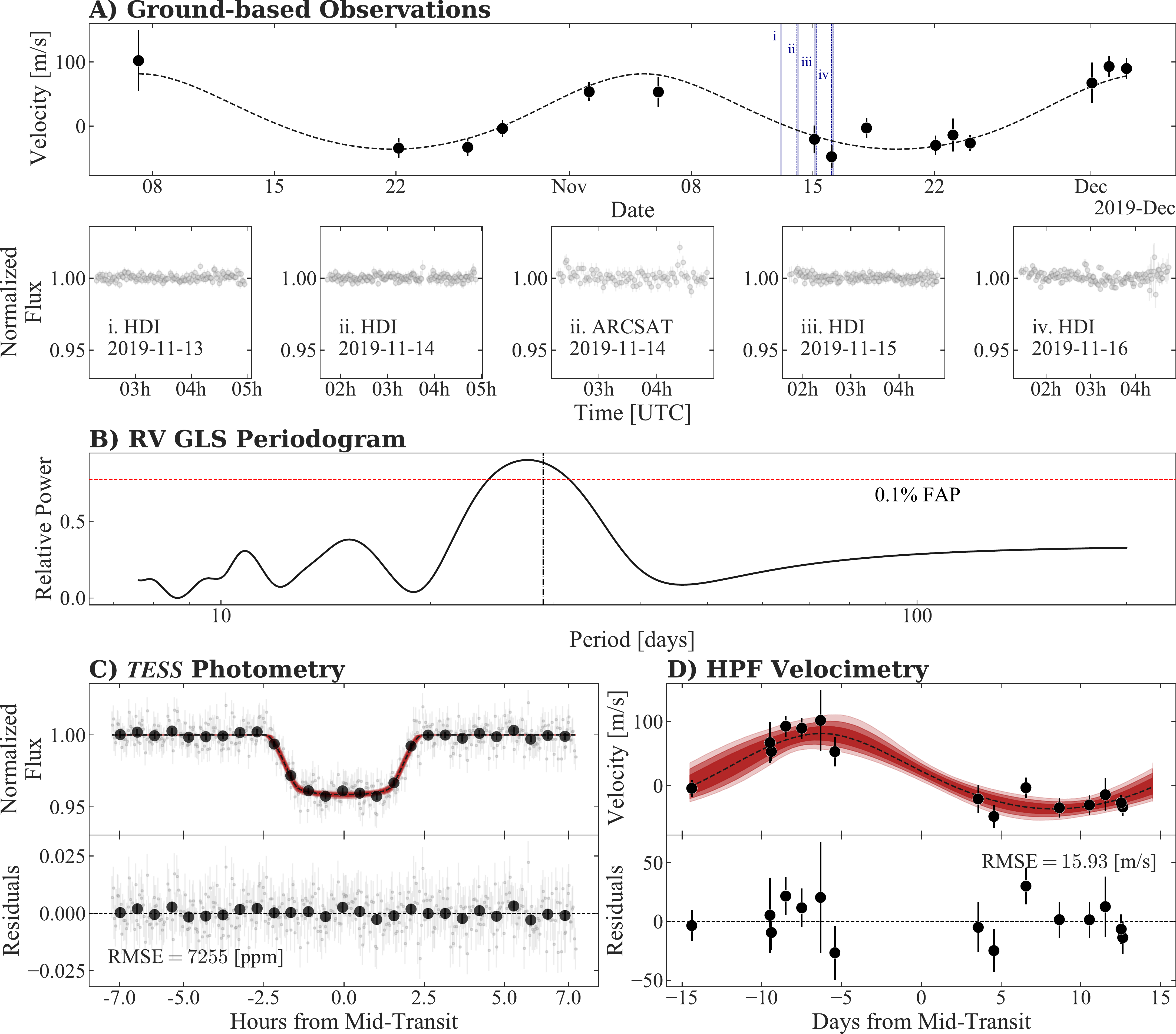}
\caption{Velocimetry and Photometry of \thistic{}. \textbf{Panel A} presents the RVs from Table \ref{tab:rvs} along with our best-fitting model denoted with a dashed line. The shaded regions mark the nights we obtained ground-based photometry. The second row shows the corresponding photometry from the respective instruments. For ease of comparison, the photometric vertical scales are identical to that of Panel B. No additional transits were detected on these nights. \textbf{Panel B} presents the generalized Lomb-Scargle periodogram of the RVs. The period from our joint fit is indicated by the vertical line. The false alarm probability (FAP) of 0.1\% is shown with the horizontal line. \textbf{Panel C} presents the \tess{} photometry around the single-transit event and \textbf{Panel D} contains the phase-folded HPF RVs. In each case, the best-fitting model is plotted as a dashed line while the shaded regions denote the \(1\sigma\) (darkest), \(2\sigma\), and \(3\sigma\) range of the derived posterior solution.}
\label{fig:rvs}
\end{figure*}

Table \ref{tab:table2} provides a summary of the inferred system parameters and respective confidence intervals. The uncertainties from the model-dependent stellar parameters are analytically propagated when calculating the values of the parameters \(M_{p}\), \(R_{p}\), \(\rho_{p}\), \(T_{\mathrm{eq}}\), \(\langle F \rangle\), and \(a\). The data reveal a companion having a mass of $0.66\pm0.07\ \mathrm{M_{J}}$ and a radius of $1.15_{-0.05}^{+0.04}\ \mathrm{R_{J}}$ transiting \thistic{} on a \(29.02_{-0.23}^{+0.35}\) day orbit. The majority of the uncertainty (\(>50\%\) of the 1\(\sigma\) confidence intervals) in the mass and radius measurements is due to the quality of the existing observations such that these measurements can be improved with photometry and RVs from more precise instruments.

Given the sparsity of the HPF data, we looked at the generalized Lomb-Scargle (GLS) periodogram \citep{Zechmeister2009} of the RVs to determine if this period solution was unique. The GLS periodogram is shown in Figure \hyperref[fig:rvs]{3B} with our best-fit period denoted by a vertical line. The RV data only show the existence of orbits near this period, as no other peaks are above a false-alarm probability (FAP) of 0.1\%. Panels (c) and (d) of Figure \ref{fig:rvs} present the result of our joint fit to the photometry and velocimetry. 

\startlongtable
\begin{deluxetable*}{llc}
\tablecaption{Derived Parameters for the \thistic{} System \label{tab:table2}}
\tablehead{\colhead{~~~Parameter} &
\colhead{Units} &
\colhead{Value}
}
\startdata
\sidehead{Photometric Parameters:}
~~~Linear Limb-darkening Coefficient\dotfill & $u_1$\dotfill & 
$0.14_{-0.10}^{+0.17}$ \\
~~~Quadratic Limb-darkening Coefficient\dotfill & $u_2$\dotfill & 
$0.22_{-0.23}^{+0.35}$ \\
\sidehead{Orbital Parameters:}
~~~Orbital Period\dotfill & $P$ (days) \dotfill & $29.02_{-0.23}^{+0.36}$\\
~~~Time of Periastron\dotfill & $T_P$ (BJD\textsubscript{TDB})\dotfill & $2458705.37_{-2.48}^{+2.28}$\\
~~~Eccentricity\dotfill & $e$ \dotfill & $0.118_{-0.077}^{+0.073}$\\
~~~Argument of Periastron\dotfill & $
\omega$ (degrees) \dotfill & $-13_{-28}^{+27}$\\
~~~Semi-amplitude Velocity\dotfill & $K$ (m/s)\dotfill &
$59.91_{-6.32}^{+6.41}$\\
~~~HPF RV Offset \dotfill & $\gamma_{HPF}$ (m/s)\dotfill & 
$16.64_{-5.23}^{+5.39}$\\
~~~RV Jitter\dotfill & $\sigma_{HPF}$ (m/s)\dotfill & $0.39_{-0.36}^{+3.84}$\\
\sidehead{Transit Parameters:}
~~~Time of Conjunction\dotfill & $T_C$ (BJD\textsubscript{TDB})\dotfill & $2458711.957792_{-0.001179}^{+0.001182}$\\
~~~Scaled Radius\dotfill & $R_{p}/R_{*}$ \dotfill & 
$0.194_{-0.005}^{+0.004}$\\
~~~Scaled Semi-major Axis\dotfill & $a/R_{*}$ \dotfill &
$56.22_{-1.66}^{+1.59}$\\
~~~Orbital Inclination\dotfill & $i$ (degrees)\dotfill &
$89.77_{-0.14}^{+0.15}$\\
~~~Impact Parameter\dotfill & $b$\dotfill &
$0.22_{-0.14}^{+0.15}$\\
~~~Transit Duration\dotfill & $T_{14}$ (hours)\dotfill &
$4.67_{-0.10}^{+0.12}$\\
~~~Photometric Jitter\dotfill & $\sigma_{TESS}$ (ppm)\dotfill & $0.01_{-0.01}^{+5.62}$\\
\sidehead{Planetary Parameters:}
~~~Mass\dotfill & $M_{p}$ (\unit{M_{J}})\dotfill &  $0.66\pm0.07$\\
~~~Radius\dotfill & $R_{p}$  (\unit{R_{J}}) \dotfill&
$1.15_{-0.05}^{+0.04}$\\
~~~Density\dotfill & $\rho_{p}$ (g/\unit{cm^{3}})\dotfill & $0.54_{-0.10}^{+0.09}$\\
~~~Surface Gravity\dotfill & $\log(g_{p})$ (cgs)\dotfill & $3.095_{-0.056}^{+0.053}$\\ 
~~~Semi-major Axis\dotfill & $a$ (au) \dotfill & $0.1587_{-0.0075}^{+0.0067}$\\
~~~Average Incident Flux\dotfill & $\langle F \rangle$ (\unit{10^8\ erg/s/cm^2})\dotfill & 
$0.039\pm0.003$\\
~~~Equilibrium Temperature\(^{a}\)\dotfill & $T_{eq}$ (K)\dotfill & 
$362\pm7$\\
\enddata
\tablenotetext{a}{The planet is assumed to be a black body.}
\normalsize
\end{deluxetable*}

\section{Discussion}
\label{sec:discussion}

\begin{figure*}[!ht]
\epsscale{1.15}
\plotone{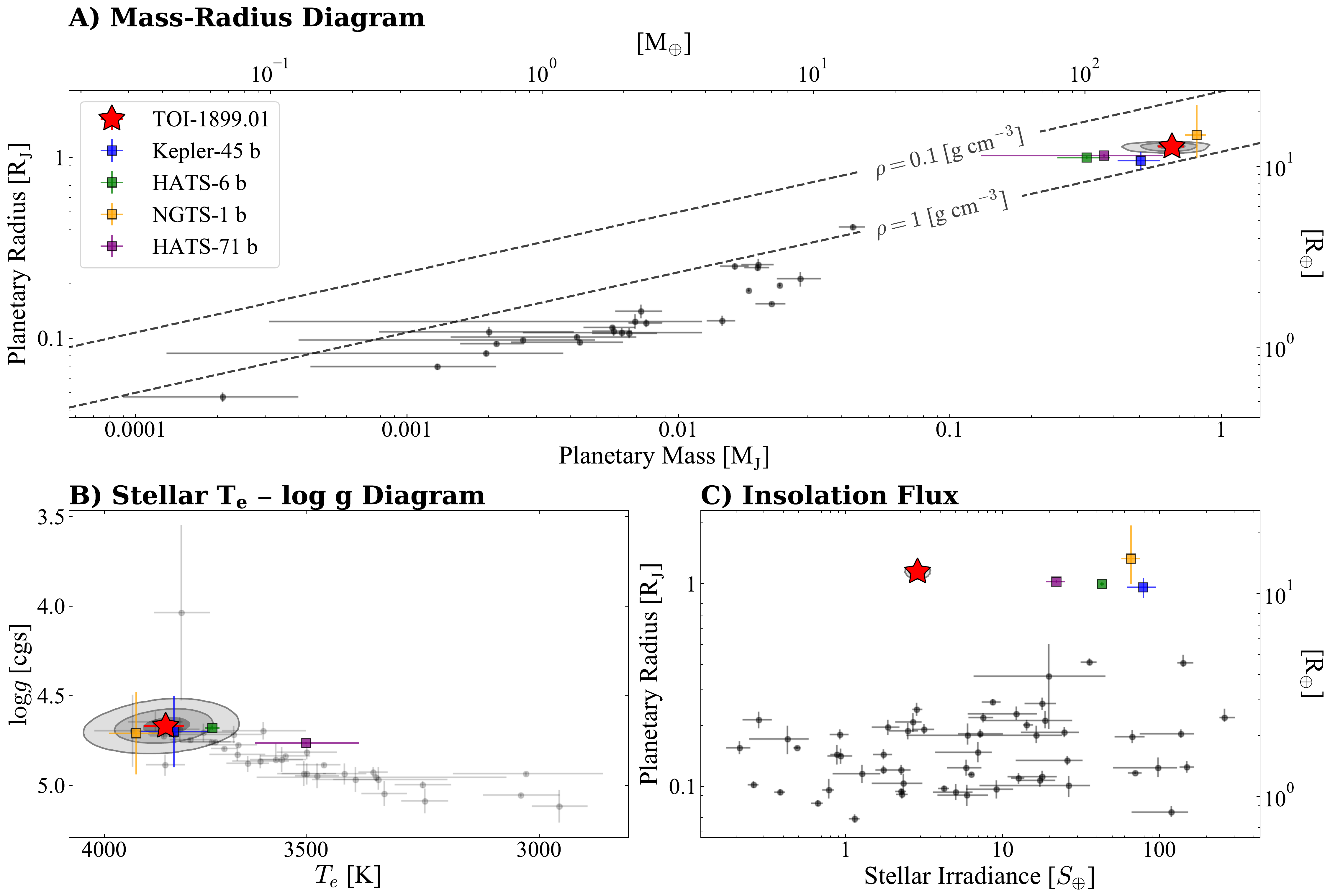}
\caption{Physical parameters of the M dwarf \thistic{} and its WJ. \textbf{Panel (a)} places the WJ, \thistic{}.01, on the mass-radius diagram for all characterized M dwarf exoplanets. For comparison, known hot Jupiters are labeled. \textbf{Panel (b)} highlights the position of \thistic{} along with other M dwarf-hosting hot Jupiters on a effective temperature - surface gravity diagram. In each of Panels (a) and (b), the posterior distribution for the relevant body of \thistic{} is shown. \textbf{Panel (c)} presents the stellar irradiation for the M dwarf exoplanets. Hot Jupiters around M dwarfs are highlighted for comparison. In each panel, the \(1\sigma\), \(2\sigma\), and \(3\sigma\) contours for the posterior distribution are shown for reference.  The data were compiled from the \href{https://exoplanetarchive.ipac.caltech.edu/cgi-bin/TblView/nph-tblView?app=ExoTbls&config=planets}{NASA Exoplanet Archive} on 2020 July 10.} 
\label{fig:exo}
\end{figure*}

\subsection{Stellar Density Diagnostic}
We used the stellar density obtained from fitting the SED as a confirmation that the transit occurs on the M dwarf \thistic{} and not the giant \wrongtic{}. The density diagnostic, in which the density derived from a transit is compared to a separate density estimate derived from stellar models, was first described by \cite{Seager2003} and has been used to examine the planetary nature of candidate planets from \kep{} and CoRoT \citep[e.g.,][]{Tingley2011}. \gaia{} DR2 provides a robust constraint on the density of a host star given the parallax and observed photometric magnitudes. 

The joint fit includes a prior on the stellar density. As an additional test, we separately fit the \tess{} photometry and HPF RVs with no density prior. The stellar density derived from the transit with no prior is \(\rho_{*,\mathrm{transit}}=3.05_{-1.32}^{+1.35}\) g/\unit{cm^{3}}, while the model-dependent density listed in Table \ref{tab:stellarparam} is \(\rho_{*,\mathrm{MIST}}=3.97_{-0.30}^{+0.37}\) g/\unit{cm^{3}}. These values agree to within $1\sigma$ and are very different from the density of \wrongtic{}, \(\rho_{*}=0.017^{+0.005}_{-0.004}\) g/\unit{cm^{3}}.

\subsection{Implications for Planetary Formation}
The WJ \thistic{}.01 is the first transiting WJ orbiting an M dwarf and only the fifth M dwarf system with a transiting Jupiter-sized planet (see Figure \ref{fig:exo}). Studies from RV surveys have shown that most low-eccentricity WJs lack giant planet companions with periods less than a few hundred days \citep{Dong2014,Bryan2016} and that metal-poor stars preferentially host low-eccentricity WJs; in contrast, metal-rich star WJs have a range of eccentricities \citep{Dawson2013}. An analysis of the \kep{} mission \citep{Huang2016} revealed that \kep{} hot Jupiters rarely have detectable inner or outer planetary companions, while half of the \kep{} WJs have close, small planetary companions. \cite{Huang2016} postulated that WJs with close planetary companions should have low orbital eccentricities and mutual inclinations, perhaps forming \textit{in-situ}, as theories where WJs form at larger distances and migrate inward (e.g., high-eccentricity tidal migration) result in the scattering of these observed companions. The existence of different populations and formation channels of WJs may be required to fully account for the properties we observe in low- and high-eccentricity WJ systems \citep{Dawson2018}.

The object \thistic{}.01 is a low-eccentricity (\(e=0.114_{-0.076}^{+0.074}\)) WJ orbiting a metal-rich star that the current data suggest lacks close massive planetary companions. It was observed by \tess{} for a total baseline of 49.9 days. The transit occurs in the middle of this window, and no additional transits or occultations were detected within the data. Our HPF RVs span a total of 56.9 days, and, to determine if the HPF data favored a long-term trend, we jointly modeled the data and included a linear trend. The resulting slope was \(\dot\gamma=0.001\pm0.013\) \unit{(mm/s)/day}, a value well below the sensitivity of HPF that provides evidence that a model with no trend is favored. The lack of additional eclipses and distortions to the standard Keplerian RV curve reveals the lack of an interior (P<29 days) massive planetary companion. However, \thistic{} could have additional exoplanets that remain undetected due to their low mass, high inclination, or long orbital periods. Additional photometric and spectroscopic observations are required to further constrain the existence of additional planetary companions.

The measurement of the apparent obliquity through the Rossiter-McLaughlin (RM) effect \citep{Triaud2018} could provide insight as to how this system formed. A direct measurement of the alignment with the host star via the RM effect would limit the physical processes involved during formation, as some mechanisms, such as disk migration, prohibit high obliquity and misalignment. The \tess{} photometry shows no activity-induced photometric variability, and a direct measurement of the stellar \(v\sin i_{*}\) is formally below the resolution of our HPF spectra. 

The large depth of this transit could make a direct measurement of the RM effect feasible. As a first-order estimate, if we assume the stellar rotation period is \(\sim\) 30 days for a well-aligned star (\(\sin i_{*}=1\)), then \(v\sin i_{*}=1\) \unit{km/s} and the expected RM effect amplitude is on the order of \(\sim35\) \unit{m/s}. While this requires a refined ephemeris, it is within the sensitivity of current precision instruments. The host star is an early M dwarf, and, given the distribution of flux and information content \citep{Reiners2018}, it is not as well suited to observation with an NIR instrument when compared to an optical or red-optical instrument. A high-precision optical instrument, such as HARPS-N \citep{Cosentino2012}, HIRES \citep{Vogt1994}, or CARMENES \citep{Quirrenbach2014,Quirrenbach2018}, would be ideal for a direct RM effect measurement.

\subsection{Implication for Planetary Interiors and Atmospheres}
\begin{figure*}[!ht]
\epsscale{1.15}
\plotone{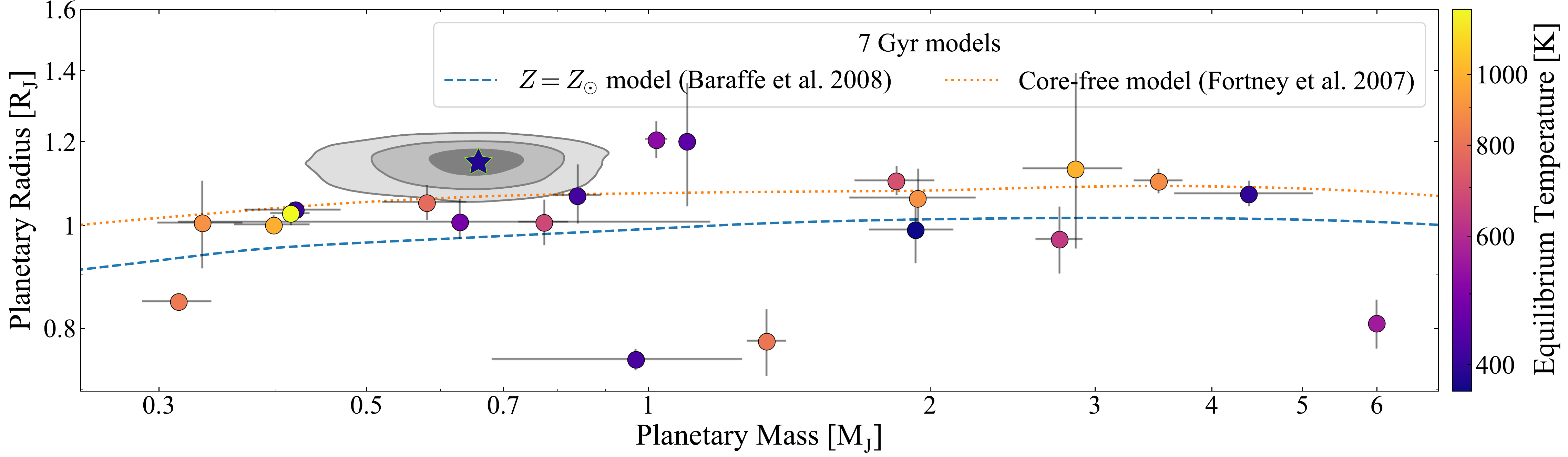}
\caption{\thistic{}.01 compared to transiting WJs. This panel presents the mass-radius diagram for transiting WJs. The \(1\sigma\), \(2\sigma\), and \(3\sigma\) posteriors for \thistic{} from our fit are included for reference. We include models from \cite{Baraffe2008} for Jovian planets at various ages without the effects of stellar irradiation. \thistic{}.01 is slightly larger than WJs of comparable mass and deviates by \(>3\sigma\) from the model from \cite{Baraffe2008} for a 7 Gyr system. The radius is consistent within \(3\sigma\) with a core-free model of a 7 Gyr system from \cite{Fortney2007} such that the large radius may result from a low solid content. It is one of the largest WJs and, given one of the lowest equilibrium temperatures, the mechanism for this inflation cannot be stellar flux-driven. The data were compiled from the \href{https://exoplanetarchive.ipac.caltech.edu/cgi-bin/TblView/nph-tblView?app=ExoTbls&config=planets}{NASA Exoplanet Archive} on 2020 July 10.} 
\label{fig:wjs}
\end{figure*}

The WJ \thistic{} has a large radius when compared to other well-characterized transiting WJs of similar mass (see Figure \ref{fig:wjs}). We compare the observed WJ radii to the radius predicted from models by \cite{Baraffe2008} of a gas giant with a solar mixture of H, He, and heavy elements. These models are for nonirradiated planets at varying ages. The observed WJs typically have radii that are within \(1\sigma\) of the predicted values, with the exception of \thistic{} and Kepler-87 b \citep{Ofir2014}, both of which deviate \(>3\sigma\) from tracks of comparable ages (\(>1\) Gyr). The radius of \thistic{}.01 may be the result of a very solid-poor composition. As an extreme case, we compare core-free models from \cite{Fortney2007} and find that the planet's radius is consistent within \(3\sigma\) of that model.

Alternatively, the planet could have an inflated radius; however, stellar flux-driven mechanisms are unlikely to be the cause. \cite{Demory2011} used a sample of giants in \kep{} to determine that gas giants receiving an incident flux \(\lesssim 2 \times 10 ^{8}\) \unit{erg/s/cm^{2}} have radii that are independent of the stellar incident flux. This threshold flux roughly corresponds to an equilibrium temperature for which ohmic heating \citep{Batygin2011} is thought to become important in heating the inner layers of a gas giant. The WJ \thistic{}.01 receives an average flux of \(0.039\times 10^{8}\) \unit{erg/s/cm^{2}}, a value well below this limit.

One possible mechanism that could result in the inflated radius despite the low stellar irradiation is delayed contraction. \cite{Baraffe2014} described two variations of delayed contraction due to an enhancement in atmospheric opacities \citep{Burrows2007} or a reduction in the interior heat transport of a planet \citep{Chabrier2007}. \cite{Burrows2007} suggested that an atmosphere with enhanced opacities (e.g., through enhanced atmospheric metallicity) would slow the cooling of a planet and maintain a larger radius for longer periods of time. This may not be an effective method of inflation, as a larger opacity through enrichment of the atmosphere requires an increased molecular weight, which may result in a smaller radius in the absence of extensive stellar irradiation \citep[e.g.,][]{Guillot2005,Guillot2008}. The second variation of delayed contraction was suggested by \cite{Chabrier2007} where the presence of a gradient of heavy elements can decrease the heat transport efficiency and slow down planetary cooling and contraction. A gradient in the mean molecular weight can prevent large-scale convection, disrupting heat transport and resulting in a semiconvective layer independent of stellar incident flux. Additional photometric observations of \thistic{}.01 are required to identify its atmospheric properties and composition and determine if the atmosphere is enriched or if nonobservable chemical gradients must be considered to inflate the radius of \thistic{}.01. 

The large radius suggests that \thistic{}.01 has a large atmospheric scale height and, potentially, large transmission spectral signals. It is cool enough that we expect the presence of molecular clouds \citep[e.g.,][]{Burrows1999,Morley2014}. \cite{Sing2016} demonstrated that while it is possible to detect the absorption feature of various molecular species in gas giants, it is difficult to predict the spectral features of a particular exoplanet given the wide range in surface gravity, metallicity, and temperatures for these objects. All of these parameters can affect a planet's atmospheric structure, circulation, and condensate formation, which in turn impact the observable features. If we assume that the composition of the atmosphere is dominated by a hydrogen-helium mixture \citep{Sing2018} and ignore the presence of clouds, we estimate absorption features with amplitudes on the order of 150 ppm. From existing atmospheric models for giant planets, we expect that the presence and height of condensates would weaken or even erase spectral features \citep[e.g.,][]{Marley1999,Sudarsky2003,Fortney2005,Morley2014}. The presence of clouds has served as a possible explanation for the weak water features of HD 209458 b \citep{Deming2013} and HAT-P-12 b \citep{Line2013} and the featureless spectra of GJ 1214 b \citep{Kreidberg2014} and GJ 436 b \citep{Knutson2014}. In the infrared, the scattering and absorption efficiencies of condensates change and can produce windows where the spectra are not significantly affected by certain clouds \citep{Morley2014}. Upcoming missions, such as the James Webb Space Telescope (JWST), will have the precision and wavelength coverage to attempt these measurements. The JWST transmission spectra of a cloudy atmosphere have the potential to constrain key model atmospheric parameters such as metallicity, C/O ratio, and various cloud parameters for cool WJs \citep{Mai2019}.

\section{Summary} \label{sec:summary}
We have confirmed the planetary nature of an object creating a single transit in a star observed by  \tess{}. The object \thistic{}.01 is the first WJ transiting an M dwarf in a low-eccentricity \(\sim29\)-day orbit. The available data do not provide evidence for massive interior planetary companions. In the population of well-characterized WJs, this planet stands out as an inflated, cool object. It is among the largest in radii, which may point toward a low fraction in solids or possibly inflation despite its cool temperature. The long period of \thistic{}.01 has the potential to make ground-based transit searches difficult, but it should be amenable to additional observations with space assets, such as the recently launched CHaracterizing ExOPlanet Satellite mission \citep[CHEOPS;][]{Broeg2013,Fortier2014}. CHEOPS has the potential to detect an additional transit for a significant fraction \citep[\(\sim70\%\);][]{Cooke2020} of single-transiting objects, such as \thistic{}, that were observed during the primary \tess{} mission. Future observations that can provide information on the atmospheric properties or formation pathways, such as atmospheric characterization or a stellar obliquity measurement, are dependent on a more precise ephemeris. We urge the community to observe this system with additional RV observations as well as for additional transits to precisely determine the period and refine constraints on the eccentricity.

\acknowledgments
% Grants
We thank the anonymous referee for a thoughtful reading of the manuscript and comments that improved the quality of this publication. CIC and GKS acknowledge support by NASA Headquarters under the NASA Earth and Space Science Fellowship Program through grants 80NSSC18K1114 and NNX16AO28H, respectively. CIC acknowledges support by the Alfred P. Sloan Foundation’s Minority Ph.D. Program under grant G-2016-20166039. GKS is also supported by the Henry Norris Russell Fellowship at Princeton University. HML acknowledges support from NSF grant AST 1616636. RID acknowledges support from grant NNX16AB50G awarded by the NASA Exoplanets Research Program and the Alfred P. Sloan Foundation's Sloan Research Fellowship.

% CEHW
This work was partially supported by funding from the Center for Exoplanets and Habitable Worlds (CEHW). CEHW is supported by the Pennsylvania State University, the Eberly College of Science, and the Pennsylvania Space Grant Consortium.

% HET/HPF, also needs footnote
This is University of Texas Center for Planetary Systems Habitability Contribution 0004. These results are based on observations obtained with the Habitable-zone Planet Finder Spectrograph on the HET. We acknowledge support from NSF grants AST 1006676, AST 1126413, AST 1310875, and AST 1310885 and the NASA Astrobiology Institute (NNA09DA76A) in our pursuit of precision radial velocities in the NIR. We acknowledge support from the Heising-Simons Foundation via grant 2017-0494.  The Hobby-Eberly Telescope is a joint project of the University of Texas at Austin, the Pennsylvania State University, Ludwig-Maximilians-Universität München, and Georg-August Universität Gottingen. The HET is named in honor of its principal benefactors, William P. Hobby and Robert E. Eberly. The HET collaboration acknowledges the support and resources from the Texas Advanced Computing Center. We thank the Resident astronomers and Telescope Operators at the HET for the skillful execution of our observations with HPF. 

% Diffuser Grant (AO)
We acknowledge support from NSF grant AST-1909506 and the Research Corporation for precision photometric observations with diffuser-assisted photometry.

% JPL (for Sam)
Part of this research was carried out at the Jet Propulsion Laboratory, California Institute of Technology, under a contract with the National Aeronautics and Space Administration (NASA).

% ACI & Cyberlamp
Computations for this research were performed on the Pennsylvania State University’s Institute for Computational and Data Sciences Advanced CyberInfrastructure (ICDS-ACI), including the CyberLAMP cluster supported by NSF grant MRI-1626251.

% ShaneAO
These results are based on observations obtained with the 3\,m Shane Telescope at Lick Observatory. The authors thank the Shane telescope operators, AO operators, and laser operators for their assistance in obtaining these data.

% NESSI
Some of the observations in this paper made use of the NN-EXPLORE Exoplanet and Stellar Speckle Imager (NESSI). NESSI was funded by the NASA Exoplanet Exploration Program and the NASA Ames Research Center. NESSI was built at the Ames Research Center by Steve B. Howell, Nic Scott, Elliott P. Horch, and Emmett Quigley.

% ARCSAT
These results are based on observations obtained with Apache Point Observatory's 0.5 m ARCSAT.

% MAST
Some of the data presented in this paper were obtained from MAST. Support for MAST for non-HST data is provided by the NASA Office of Space Science via grant NNX09AF08G and by other grants and contracts.
% Kepler/TESS
This work includes data collected by the \tess{} mission that are publicly available from MAST. Funding for the \tess{} mission is provided by the NASA Science Mission directorate. 
% NASA Exoplanet Archive
This research made use of the NASA Exoplanet Archive, which is operated by Caltech, under contract with NASA under the Exoplanet Exploration Program.
% 2MASS 
This work includes data from 2MASS, which is a joint project of the University of Massachusetts and IPAC at Caltech funded by NASA and the NSF.

% APASS
We acknowledge with thanks the variable star observations from the AAVSO International Database contributed by observers worldwide and used in this research.

% Gaia
This work has made use of data from the European Space Agency (ESA) mission \gaia{} (\url{https://www.cosmos.esa.int/gaia}), processed by the \gaia{} Data Processing and Analysis Consortium (DPAC, \url{https://www.cosmos.esa.int/web/gaia/dpac/consortium}). Funding for the DPAC has been provided by national institutions, in particular the institutions participating in the \gaia{} Multilateral Agreement.

% ZTF
Some observations were obtained with the Samuel Oschin 48 inch Telescope at the Palomar Observatory as part of the ZTF project. The ZTF is supported by the NSF under grant No. AST-1440341 and a collaboration including Caltech, IPAC, the Weizmann Institute for Science, the Oskar Klein Center at Stockholm University, the University of Maryland, the University of Washington, Deutsches Elektronen-Synchrotron and Humboldt University, Los Alamos National Laboratories, the TANGO Consortium of Taiwan, the University of Wisconsin at Milwaukee, and Lawrence Berkeley National Laboratories. Operations are conducted by COO, IPAC, and UW.

% LAMOST
This work has made use of data from the Guoshoujing Telescope (LAMOST), a National Major Scientific Project built by the Chinese Academy of Sciences. Funding for the project has been provided by the National Development and Reform Commission. LAMOST is operated and managed by the National Astronomical Observatories, Chinese Academy of Sciences.

% https://journals.aas.org/facility-keywords/
\facilities{AAVSO, \gaia{}, HET (HPF), KPNO (HDI), LAMOST, PO:1.2 m (ZTF), Shane (AO), \tess{}, WIYN (NESSI)} 
\software{AstroImageJ \citep{Collins2017}, 
\texttt{astroquery} \citep{Ginsburg2019},
\texttt{astropy} \citep{AstropyCollaboration2018},
\texttt{barycorrpy} \citep{Kanodia2018}, 
\texttt{batman} \citep{Kreidberg2015},
\texttt{celerite} \citep{Foreman-Mackey2017},
\texttt{ccdproc} \citep{Craig2017},
\texttt{dustmaps} \citep{Green2018a},
\texttt{DAVE} \citep{Kostov2019},
\texttt{EXOFASTv2} \citep{Eastman2019},
\texttt{HxRGproc} \citep{Ninan2018},
\texttt{GNU Parallel} \citep{Tange2011},
\texttt{juliet} \citep{Espinoza2019},
\texttt{lightkurve} \citep{LightkurveCollaboration2018},
\texttt{matplotlib} \citep{hunter2007},
\texttt{MultiNest} \citep{Feroz2009,Feroz2013},
\texttt{numpy} \citep{vanderwalt2011},
\texttt{pandas} \citep{McKinney2010},
\texttt{Photutils} \citep{Bradley2019},
\texttt{radvel} \citep{Fulton2018},
\texttt{scipy} \citep{Virtanen2019},
\texttt{SERVAL},
\texttt{SpecMatch-Emp},
\texttt{VESPA} \citep{Morton2012}}

\end{document}